\documentstyle[aps,epsf]{revtex} 

\begin{document} 
  
\centerline{\large\bf Topological Susceptibility of Yang-Mills
Center Projection Vortices}

\vspace{0.7cm}

\centerline{R.~Bertle$^{a}$, M.~Engelhardt$^{b}$ and M.~Faber$^{a}$ }
\vspace{0.6cm}
\centerline{\em $^{a}$Atominstitut der \"osterreichischen Universit\"aten,
Arbeitsgruppe Kernphysik, TU Wien, Austria}
\vspace{0.2cm}
\centerline{\em $^{b}$Institut f\"ur Theoretische Physik,
Universit\"at T\"ubingen, Germany}
\vspace{0.6cm}

\begin{abstract}
The topological susceptibility induced by center projection vortices
extracted from $SU(2)$ lattice Yang-Mills configurations via the maximal
center gauge is measured. Two different smoothing procedures, designed to 
eliminate spurious ultraviolet fluctuations of these vortices before
evaluating the topological charge, are explored. They result in
consistent estimates of the topological susceptibility carried by the
physical thick vortices characterizing the Yang-Mills vacuum in the
vortex picture. This susceptibility is comparable to the one obtained
from the full lattice Yang-Mills configurations. The topological properties
of the $SU(2)$ Yang-Mills vacuum can thus be accounted for in terms
of its vortex content.
\end{abstract}


\medskip

\section{Introduction}
The center vortex picture of the Yang-Mills vacuum has recently 
experienced rapid development. Two advances figure prominently in 
particular. On the one hand, methods have been constructed which allow
to extract vortices from lattice gauge configurations; on the other hand,
it has been realized that the vortex picture not only can account for
confinement, but also for the topological properties of the Yang-Mills
field.

In more detail, center vortices can be extracted from lattice gauge
configurations using a combined gauge fixing and projection procedure.
The first such procedure was presented in 
\cite{deb97},\cite{deb97aug},\cite{giedt}; it involves fixing the gauge 
up to transformations from the center of the gauge group, by demanding 
maximization of a gauge fixing functional,
\begin{equation}
\max_{G} \sum_{i} \left| \mbox{tr} \, U_i^G \right|^{2}
\label{gfunct}
\end{equation}
under gauge transformations $G$, where the $U_i $ are the link variables
specifying a lattice gauge configuration. This is called the (direct)
maximal center gauge. It biases link variables towards the center of
the gauge group, e.g. in the $SU(2)$ case considered henceforth,
towards the elements $\pm 1$. Physically, the idea is to transform as much 
physical information as possible to the center part of the configuration,
i.e. the part one obtains by subsequently projecting
\begin{equation}
U \longrightarrow \mbox{sign tr} \, U
\label{cproj}
\end{equation}
This projection step truncates physical information, and whether one has
succeeded in retaining the relevant physics can usually only be
answered a posteriori, by comparing the results obtained for a 
particular observable using either the center projected configurations
or the full ones. If the results agree, this is called {\em center
dominance}.

The observation of center dominance for large Wilson loops\footnote{This 
observation was subsequently extended to finite temperatures 
\cite{temp},\cite{tlang},\cite{bertle}.} 
\cite{deb97},\cite{deb97aug},\cite{giedt} sparked the recent renewed 
interest in the {\em center vortex picture} of confinement. Once one has 
defined a $Z(2)$ lattice configuration by the center projection step 
(\ref{cproj}), one equivalently obtains an associated center vortex 
configuration in a canonical fashion: One considers all plaquettes of 
the $Z(2)$ lattice, and if the product of the links bounding a plaquette 
yields the value $-1$, a vortex is defined to pierce that plaquette. 
These vortices form closed two-dimensional world-sheets in 
four-dimensional (Euclidean) space-time, and they can be 
thought of as being composed of plaquettes on the dual lattice, i.e. the 
lattice shifted by half a lattice spacing in all four directions with 
respect to the original one. Vortices contribute a phase factor $-1$ to 
the value of any Wilson loop they are linked to. Center dominance 
equivalently implies vortex dominance.

Depending on the observable under consideration, the original or the
dual lattice (vortex) language may be advantageous. Note that, while
the formulation in terms of $Z(2)$ link variables still has a residual
$Z(2)$ gauge invariance (the gauge fixing functional (\ref{gfunct})
contains no bias with respect to transformations $G\in Z(2)$), the
vortex world-sheets on the other hand represent gauge-invariant
variables under this residual gauge group. One advantage of the vortex
language lies in the fact that it can be detached from an underlying
space-time lattice; one can consider vortex world-surfaces
in continuous space-time. This has proven particularly useful when 
considering the topological properties of vortex configurations, which
initially are only defined in a continuum framework 
\cite{cont},\cite{cornwalt},\cite{cornw}. The topological
winding number (Pontryagin index) $Q$ of an $SU(2)$ vortex surface 
configuration $S$ can be given in terms of its (oriented) 
self-intersection number \cite{cont}
\begin{equation}
Q = -\frac{1}{16} \epsilon_{\mu \nu \alpha \beta }
\int_{S} d^2 \sigma_{\alpha \beta }
\int_{S} d^2 \sigma^{\prime }_{\mu \nu }
\delta^{4} (\bar{x} (\sigma ) - \bar{x} (\sigma^{\prime } ) )
\label{yidres}
\end{equation}
where $\bar{x} (\sigma )$ denotes a parametrization of the
two-dimensional surface $S$ in four space-time dimensions, i.e. for any
two-vector $\sigma $ from a two-dimensional parameter space, the
four-vector $\bar{x} $ gives the corresponding point on the vortex
surface in space-time. This parametrization furthermore implies an
infinitesimal surface element
\begin{equation}
d^2 \sigma_{\mu \nu } = \epsilon_{ab}
\frac{\partial \bar{x}_{\mu  } }{\partial \sigma_{a} }
\frac{\partial \bar{x}_{\nu  } }{\partial \sigma_{b} }
d^2 \sigma
\label{serfel}
\end{equation}
on the vortex surface. Note that the continuum surfaces thus must be
specified not only with a location, but also an orientation, encoded in
the sign of the surface element (\ref{serfel}). In general, vortex surfaces
consist of surface patches of alternating orientation; indeed, generic
center projection vortices in the confining phase have been shown to be
non-orientable \cite{bertle}, and therefore necessarily contain lines at 
which the orientation switches, i.e. patch boundaries. These lines can be 
associated with Abelian monopole trajectories \cite{cont}. For $Z(2)$ 
lattice vortices, by contrast, the orientation information is initially 
lost and must be reintroduced before the topological charge can be 
evaluated \cite{preptop}. Note furthermore that (\ref{yidres}) includes 
not only surface self-intersections in the usual sense (for which 
(\ref{yidres}) is normalized such as to give a contribution of modulus 
$1/2$, as shown in \cite{cont},\cite{cornw}), but all singular points of a 
surface \cite{preptop}; these are all points where the set of tangent 
vectors to the surface spans all four space-time directions.

The continuum expression (\ref{yidres}) for the topological charge
has been implemented for vortex surfaces defined on a (dual) space-time
lattice \cite{preptop}; the detailed prescription is presented in 
section \ref{chilsec}. Using this construction, a random vortex surface 
ensemble defined such as to reproduce the confinement properties of 
$SU(2)$ Yang-Mills theory \cite{selprep} was shown to simultaneously 
predict the correct topological susceptibility \cite{preptop} as measured 
in lattice Yang-Mills experiments. The topological susceptibility 
$\chi = \langle Q^2 \rangle /V$, where $V$ is the space-time volume under 
consideration, e.g. determines, via the Witten-Veneziano 
estimate \cite{witven}
\begin{equation}
m^2_{\eta^{\prime } } + m^2_{\eta } - 2m^2_K = 2N_f \chi /f^2_{\pi }
\end{equation}
the anomalously high mass of the $\eta^{\prime } $ meson. These developments 
have given rise to the expectation that the vortex picture may be the first 
infrared effective framework capable of providing a consistent explanation 
of the entire spectrum of nonperturbative effects characterizing the 
infrared sector of Yang-Mills theory. Originally, the vortex picture was 
only proposed in particular to explain confinement \cite{thoo}. It received 
early corroboration through the observation that a constant Yang-Mills 
chromomagnetic field is unstable with respect to the formation of flux 
tube domains in three-dimensional space \cite{spag}. Also in the lattice 
formulation, different possibilities of defining vortices were explored
\cite{mack},\cite{tomold},\cite{hartte}. Notably, an alternative 
confinement criterion based on the vortex free energy was established 
\cite{mack} and measurements of this free energy have recently been carried 
out \cite{tomfen},\cite{tepfen}. The new developments highlighted further 
above have significantly enlarged the scope of the vortex picture beyond 
this by establishing the link to the topological properties of Yang-Mills 
theory.

The purpose of the present work is to combine the two recent advances
discussed above: The construction of the topological charge for lattice
vortex surfaces \cite{preptop}, hitherto only applied to an ensemble of 
random vortex surfaces \cite{preptop}, will be used to measure the topological 
susceptibility associated with the vortices extracted from lattice Yang-Mills
configurations via the maximal center gauge\footnote{Note that up to now
only the complementary experiment has been performed \cite{forc1}: An
ensemble of Yang-Mills configurations from which all vortices had been
removed was shown to belong exclusively to the topologically trivial
sector $Q=0$.}. These vortex configurations exhibit ultraviolet artefacts
which have been noted before in other contexts, cf. the discussion in
section \ref{eliart}. These artefacts strongly influence the measurement 
of the topological charge, and it is consequently necessary to explore
smoothing procedures applied to the vortices before the measurement.
In this respect, the center projection framework is quite similar to
full lattice Yang-Mills experiments. While the prediction for the 
topological susceptibility arrived at in this work is thus fraught 
with a measure of systematic uncertainty, the values obtained will be
seen to be comparable with the full $SU(2)$ Yang-Mills topological 
susceptibility. This result is consistent with an interpretation of the 
topological charge of Yang-Mills configurations being generated by their 
vortex content.

Finally, a comment is in order regarding the gauge chosen to extract the
vortices. The maximal center gauge to be used in the following has been
the subject of recent debate due to pronounced Gribov copy effects.
While most Gribov copies, i.e. local maxima of the gauge fixing functional
(\ref{gfunct}), appear, as an ensemble, to generate center dominance for
large Wilson loops, closer numerical scrutiny 
\cite{tomins},\cite{borny1},\cite{versbo},\cite{borny2} has
indicated that the highest maxima may not. This actually is not entirely
surprising in view of the continuum limit of the maximal center gauge
\cite{cont}, which always leads to a trivial center projection, with no 
vortex content. One remedy suggested in \cite{cont} is to consider altered 
functionals,
\begin{equation}
\max_{G} \sum_{i} f \left( \left| \mbox{tr} \, U_i^G \right| \right)
\label{cosug}
\end{equation}
with monotonously rising $f$; any such functional implements the general
idea of concentrating physical information onto the center of the gauge
group. It is in fact quite straightforward to construct functions $f$
which, in the continuum, avoid the singularities leading to the absence
of vortices in the case of the gauge fixing functional (\ref{gfunct}).
Another alternative type of gauge which allows to identify vortices
is the Laplacian center gauge 
\cite{forc2},\cite{forc3},\cite{pepe},\cite{torsten}, which is free of 
Gribov copies. A further possibility of circumventing the Gribov problem 
lies in reinterpreting the maximal center gauge fixing procedure used on 
the lattice. Instead of insisting on convergence to the global maximum
of (\ref{gfunct}), which in practice is rarely achieved, one can allow
for an averaging over different local maxima. Formally, this is akin
to the well-known procedure of introducing a gauge-fixing term with a
finite gauge fixing parameter into the action, without insisting on
any particular limit for this parameter \cite{cont},\cite{quantum}.

While it is ultimately desirable to develop better-defined alternative
gauge fixing procedures such as highlighted above, these procedures
are just beginning to be explored\footnote{For some first results using the
Laplacian center gauge, cf. 
\cite{forc2},\cite{forc3},\cite{pepe},\cite{kurt}; note e.g. the as
yet unclear interpretation of the vortex density in the Laplacian center
gauge, which does not seem to scale properly with the renormalization
group and seems to extrapolate to an infinite continuum value \cite{kurt}.}.
For this reason, in this work, the usual maximal center gauge overrelaxation
algorithm \cite{deb97},\cite{deb97aug},\cite{giedt} will be used to 
{\em define} the gauge fixing image, in recognition of the fact that it 
does not strictly correspond to the condition (\ref{gfunct}), but rather 
to some reinterpretation such as suggested above.

\section{Topological charge of lattice vortex surfaces}
\label{chilsec}
Given vortex surface configurations composed of plaquettes on a (dual)
hypercubic lattice, such as e.g. extracted via center projection
from $SU(2)$ lattice Yang-Mills configurations, the Pontryagin index
cannot immediately be evaluated. For one, as mentioned above, orientations
first have to be assigned to the plaquettes making up the surface
configuration. In the case of center projection vortices, one could
use an indirect version of the maximal center gauge 
\cite{deb97},\cite{deb97aug}, in which one first transforms to the 
maximal Abelian gauge, extracts the Abelian monopoles via Abelian 
projection, and subsequently fixes the residual Abelian U(1) gauge 
freedom such as to reach the maximal center gauge. The monopoles define 
the edges of the oriented patches making up the vortex surfaces. 
However, this gauge fixing procedure does not strictly locate the 
monopoles on the surfaces; there remains a small fraction of 
monopole links which do not lie on a vortex\footnote{In this respect,
the Laplacian center gauge 
\cite{forc2},\cite{forc3},\cite{pepe},\cite{torsten} is more
advantageous, since it can be constructed such as to explicitly locate 
monopoles on vortex surfaces.} \cite{deb97aug}. In anticipation of the fact
that the topological charge varies very little under most (topologically
allowed) deformations of the monopole trajectories (the reason for this
is discussed in detail in section \ref{ress}), in this work a simpler 
prescription was applied. Namely, vortex surfaces were identified by
transforming directly to the maximal center gauge; then, random initial 
orientations were assigned to the vortex plaquettes, and sweeps through the 
lattice were performed in which plaquettes were reoriented such as to either 
maximize or minimize the monopole density. The topological susceptibility 
was evaluated for both cases and turns out to be equal within the 
statistical error of the measurement, even though the monopole density 
varies by a large factor; numerical values are given in section \ref{ress}.

The oriented vortex surface configurations defined in this way contain
two types of ambiguities as to their precise continuum interpretation,
which must be resolved before the Pontryagin index can be evaluated. 
One ambiguity lies in the fact that the vortex
surfaces generically intersect along whole lines, as opposed to points,
as they do in the continuum. Similarly, monopole trajectories on the
surfaces (at which the surface orientation switches) intersect singular
points (at which topological charge is generated) with a finite
probability; in the continuum, assuming a random distribution of lines
and points on a surface, the two sets are generically disjoint. These
ambiguities arise due to the coarse-graining enforced by the lattice;
to resolve them, one defines the vortex configurations on a finer
lattice and allows for random small deformations of the surfaces and of
the monopoles until the ambiguities are resolved \cite{preptop}. This is 
reminiscent of inverse blocking prescriptions in full Yang-Mills theory. 
In summary, one arrives at the following algorithm for evaluating the 
Pontryagin index of the lattice vortex surfaces (the reader interested in 
further details of the construction is referred to \cite{preptop}):

\begin{itemize}
\item 
Transfer the given lattice surface configuration onto a lattice
of $1/3$ the lattice spacing and sweep once through the lattice, applying
so-called elementary cube transformations whenever this allows to remove 
one of the links along which two vortex surface segments intersect, i.e. 
to which more than two vortex plaquettes are attached. An elementary cube 
transformation\footnote{Elementary cube transformations are also the
basic building blocks of the smoothing procedure discussed in section
\ref{smosec}, albeit with a different acceptance criterion; 
Fig.~\ref{smodef} in section \ref{smosec} displays particular examples of 
elementary cube transformations used in the context of smoothing.}
is a particular elementary update of a lattice surface configuration; it is 
carried out on all six plaquettes of an elementary three-dimensional cube of 
the lattice. If any of these plaquettes were part of a vortex before the
update, then they are removed from the vortex after the update, and
vice versa. Note that an elementary cube transformation preserves the
closed character of the surfaces. The orientations of the updated
plaquettes should be chosen such as to conserve the number of monopole
links as much as possible. In practice, it is sufficient to carry this
procedure out twice, i.e. one ends up with a lattice of $1/9$ the 
original lattice spacing. As a result, the surface configuration only
self-intersects at points, but not anymore along lines.
\item 
For each lattice site $n$, make a copy of all attached plaquettes,
and transform the copy as follows. After finding an initial plaquette
which is part of a vortex, iteratively reorient further vortex plaquettes 
sharing links with previously considered ones such as to remove all 
monopole lines. If there are two independent surface segments present, i.e.
two sets of vortex plaquettes which share only the lattice site under 
consideration, then carry out this procedure for each segment independently.
The result of this procedure is that all monopole lines are deformed away
from the lattice site being scrutinized. Using the transformed vortex 
plaquettes, one can now obtain the contribution $q_n $ to the Pontryagin 
index from the site $n$ in question, cf. eq. (\ref{yidres}) and the 
discussion following. Namely, each pair of mutually orthogonal vortex 
plaquettes, i.e. with combined tangent vectors spanning four dimensions, 
contributes $\pm 1/32$ to the Pontryagin index. The sign depends on the 
relative orientation of the two plaquettes; the magnitude $1/32$ can be 
inferred from the fact that an intersection point in the usual sense 
contains $16$ such pairs, and carries topological charge $q_n = \pm 1/2$, 
cf. \cite{cont},\cite{cornw}.
\item 
The total Pontryagin index is the sum $Q=\sum_{n} q_n $.
\end{itemize}

Applying this procedure to general surface configurations on a hypercubic 
lattice, i.e. a space-time torus, one obtains a Pontryagin index quantized 
in half-integer units, which is exactly as it should be \cite{baal}; this
is nontrivial in view of the fact that the magnitudes of the individual 
site contributions $q_n $ can be as small as $1/16$ (for surfaces in a 
space-time continuum, the contribution from any particular singular point
can even be arbitrarily small).

\section{Elimination of ultraviolet artefacts}
\label{eliart}
\begin{figure}
\centerline{
\epsfxsize=6cm
\epsffile{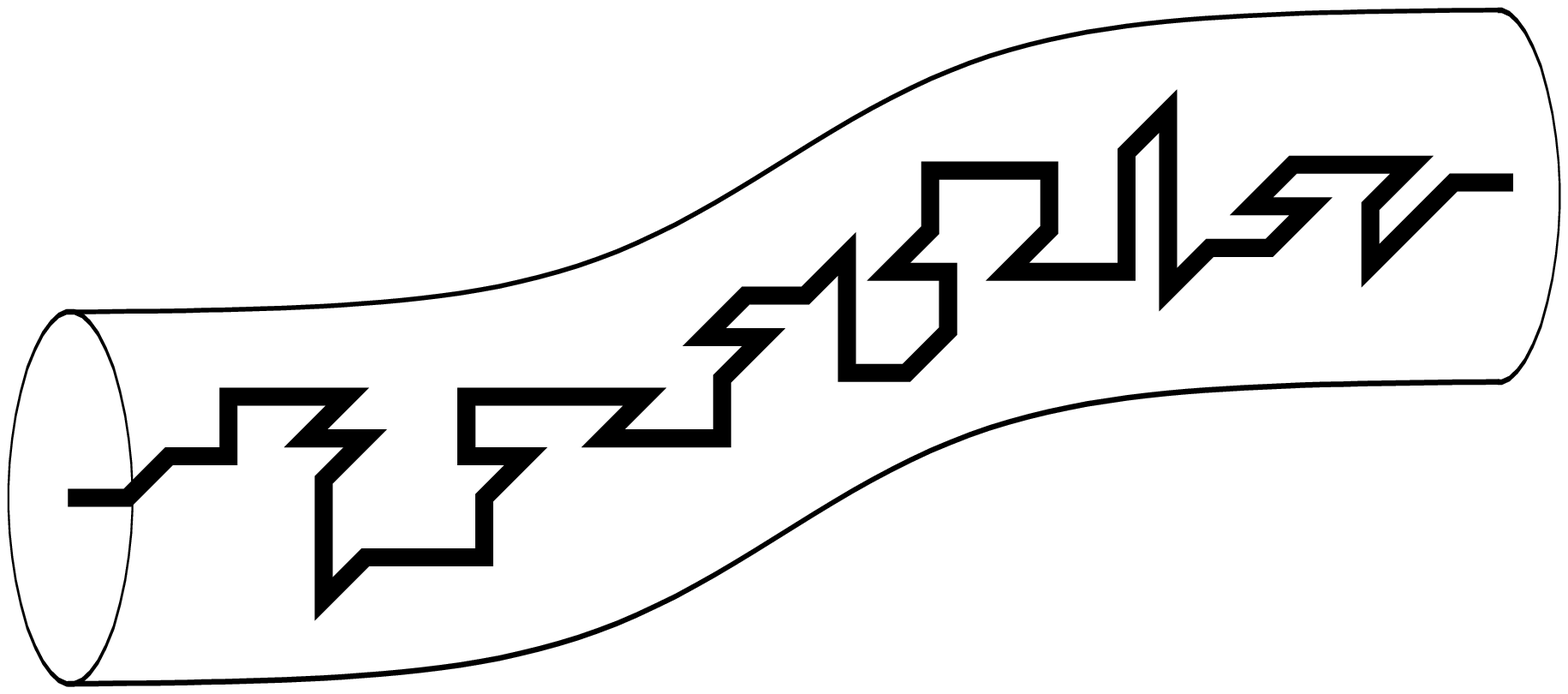}
\hspace{1.6cm}
\epsfxsize=6cm
\epsffile{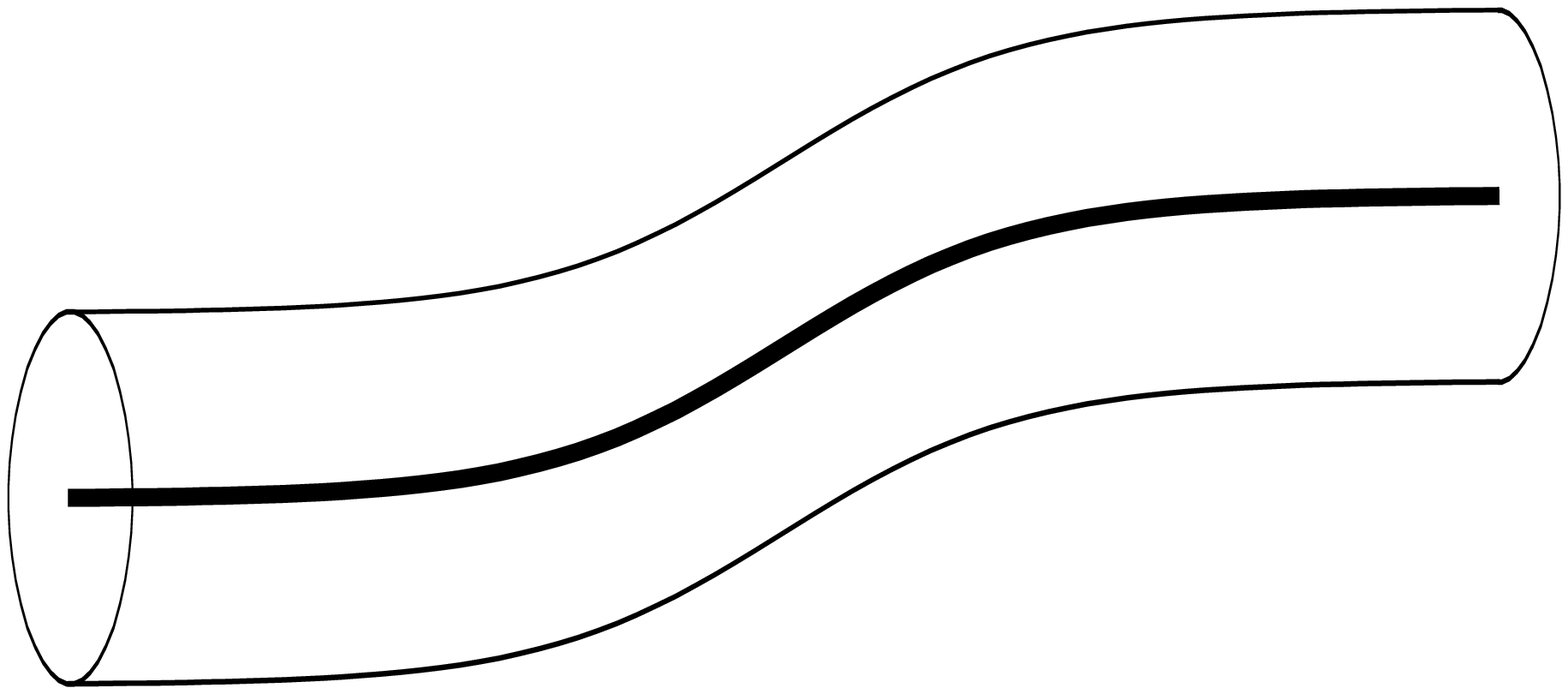}
\vspace{0.3cm}
}
\caption{\em Thick vortices (in a three-dimensional slice of space-time)
associated with rough lattice center projection vortices (left), and
smooth thin vortex cores (right), respectively.}
\label{flucpv}
\end{figure}

Using the above algorithm, the Pontryagin index of an arbitrary lattice
center projection vortex configuration can be evaluated. However, such
thin vortex surfaces exhibit ultraviolet artefacts which 
strongly contaminate the measurement. Specifically, in the vortex 
picture of Yang-Mills theory, thick physical vortices are conjectured
to characterize the infrared properties of the vacuum. By contrast,
the thin vortices obtained by center projection from full Yang-Mills 
configurations only approximate the aforementioned thick vortices
on infrared length scales coarser then a scale related to the physical 
vortex thickness. On the other hand, within the thick profile of the 
physical vortex it corresponds to, the thin center projection vortex in 
general exhibits spurious gauge-dependent ultraviolet fluctuations, cf. 
Fig. \ref{flucpv}. This has been noted before in several contexts. For
one, in \cite{cont} explicit examples were constructed showing that 
the precise location of a center projection vortex on scales finer than the 
physical vortex thickness depends on the gauge fixing function $f$ in 
(\ref{cosug}). Similarly, in \cite{giedt} it was observed that this location 
varies as different Gribov copies of the maximal center gauge fixed 
configurations are considered.

Moreover, in lattice experiments, the ratio between the string tension and 
the center projection vortex density is substantially suppressed compared 
to what one would expect from a random vortex ensemble \cite{temp}; 
motivated by this, the correlations between intersection points of center
projection vortices with a given space-time plane were investigated
in \cite{corr}. On ultraviolet length scales, up to about $0.4$ fm,
the binary correlation function between these points indeed is strongly
enhanced. Also this finding can be understood in terms of the aforementioned
short wavelength fluctuations of the center projection vortices. Consider
a plane which cuts a thick vortex, such as depicted in Fig. \ref{flucpv}, 
along a (smeared-out) line, and consider furthermore intersection points of 
the associated center projection vortex with this plane. Due to the 
transverse fluctuations of the projection vortices, one will find a
strongly enhanced probability of detecting such intersection points close 
to one another (compared with the probability one would expect from the 
mean vortex density). Note that this picture also clarifies the origin 
of the low ratio of the string tension to the center projection vortex 
density mentioned further above; the vortex density relevant for the 
area law behavior of the Wilson loop is the density of thick vortices, 
which would be well represented by smooth thin vortex cores, 
cf. Fig. \ref{flucpv}, as opposed to the rough center projection vortices.
Indeed, the smoothing procedure to be discussed in section \ref{smosec}
has been shown to keep the string tension approximately fixed while
depleting the vortex density by eliminating ultraviolet fluctuations
\cite{bertle}; ultimately, one reaches a ratio compatible with the
one expected in a random vortex ensemble. 

As a consequence of these observations, it is necessary to explore
methods to eliminate the spurious ultraviolet fluctuations of the
center projection vortices; only then can one expect to extract the 
physical topological content of the configurations, without contamination
by ultraviolet artefacts. Before discussing these methods, it is worth
noting that the problems discussed above would be exacerbated by using
the Laplacian center gauge instead of the maximal center gauge. In the
latter, the center projection vortex density and its binary correlations
measured in \cite{corr} scale properly under the renormalization group,
such as to extrapolate to a finite physical result in the continuum
limit, albeit with strong correlations in the ultraviolet, as discussed
above. By contrast, it has recently been noted that in the Laplacian
center gauge, the vortex density does not scale to a finite continuum
limit, but appears to extrapolate to an infinite continuum density 
\cite{kurt}. In a sense, the Laplacian center gauge may suffer from
its own efficiency. As observed in \cite{pepe}, vortices extracted via
the Laplacian center gauge at least partly reproduce the short-range
Coulomb potential between static charges, whereas this effect is
completely truncated when projecting from the maximal center gauge.
In terms of the underlying degrees of freedom, this presumably means
that, in addition to the infrared structure of the theory,
Laplacian center gauge fixing attempts to also partially represent 
ultraviolet perturbative gluons by vortices. This may be the
reason for the unphysical renormalization group behavior observed for
the density of Laplacian center gauge vortices \cite{kurt}.

\subsection{Blocking}
\label{blosec}
A simple way to eliminate ultraviolet fluctuations of the center projection
vortices obtained in the maximal center gauge is to apply blocking steps
such as to transfer the vortex configurations onto new coarser lattices,
while always preserving their chromomagnetic flux content on length
scales larger than the new lattice spacing. Such blocking steps are
easily implemented starting on the original lattice, i.e. before
constructing the vortex surfaces on the corresponding dual lattice.

Consider a new coarse lattice with $n$ times\footnote{In practice,
$n=2,3$ and $4$ were used.} the spacing of an old fine lattice,
superimposed on the latter such that all sites of the coarse lattice
coincide with sites of the fine lattice. The gauge phases associated
with plaquettes on the coarse lattice then are defined to be equal to
the $n\times n$ Wilson loops on the old fine lattice to which these
plaquettes correspond. Equivalently, if an odd number of vortices
pierces the $n\times n$ Wilson loop on the old fine lattice, then one
vortex is defined to pierce the corresponding plaquette on the new coarse 
lattice; if an even number of vortices pierces the $n\times n$ Wilson loop
on the fine lattice, then no vortex pierces the corresponding plaquette 
on the coarse lattice. Note that, in practice, this procedure (and also
the smoothing procedure discussed further below) was applied to the
center projected lattice configurations {\em before} defining the
orientations of the vortex surfaces in the manner described at the
beginning of section \ref{chilsec}. Note also that blocking manifestly
preserves the values of all Wilson loops (as far as they can still be
defined on the coarse lattice). Thus, blocking leaves the string tension
induced by a thin vortex ensemble invariant.

In section \ref{ress}, the behavior of the topological susceptibility as 
a function of the coarse lattice spacing reached by blocking is discussed.
Of course, the question arises which scale defines the separation between
spurious ultraviolet fluctuations to be eliminated, and relevant infrared
information on vortex degrees of freedom to be kept. Obviously, this
scale is related to the thickness of the physical vortices thought to
be present in the full Yang-Mills configurations; however, the precise
relation is a priori unclear. Certainly, the blocking procedure should not 
be carried so far as to deplete the density of thick vortices relevant for 
the asymptotic string tension. An estimate of this density, discussed in
\cite{berbang}, leads to the conclusion that the centers of neighboring
thick vortices, on the average, are $0.6$ fm apart. This therefore
constitutes an upper bound on the length scales to be eliminated by
the blocking procedure. On the other hand, the ultraviolet correlations
measured in \cite{corr}, interpreted above to be a consequence of the
spurious short wavelength fluctuations of the center projection vortex
surfaces, extend to distances up to about $0.4$ fm. Thus, the authors
estimate that the separation scale, i.e. the new lattice spacing which 
should be reached by blocking, roughly lies between $0.4$ fm and $0.6$ fm. 
This is also compatible with the findings in a random vortex surface model
\cite{selprep} adjusted to reproduce the confinement properties of
$SU(2)$ Yang-Mills theory; there, two neighboring vortices can be
identified as distinct down to a minimal distance of $0.4$ fm.

It should be emphasized that the above estimates do not exclude the 
chromomagnetic flux of the vortices being smeared out considerably further;
i.e., the flux of neighboring thick physical vortices may to a certain 
extent overlap \cite{berbang}. The flux of a physical vortex has been 
argued to extend transversally over a distance of a little over $1$ fm 
\cite{giedt},\cite{whatare}, in order to account e.g. for the Casimir 
scaling behavior of adjoint representation Wilson loops \cite{fa97}.

\subsection{Smoothing}
\label{smosec}
\begin{figure}
\centerline{
\epsfysize=7.5cm
\epsffile{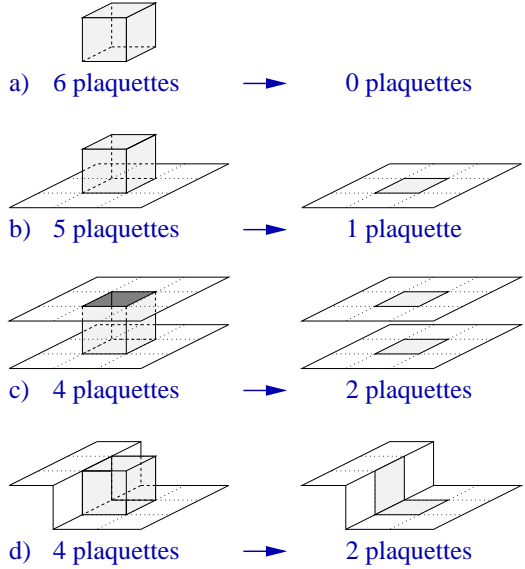}
\vspace{0.5cm}
}
\caption{\em Different smoothing steps effected by elementary cube
transformations. They are distinguished by the number of vortex plaquettes
removed and created by the operation. Note that two possibilities
c) and d) of removing four plaquettes and creating two other ones are
possible, depending on whether the latter two are opposite (c) or
adjacent faces (d) of the elementary cube. The ordering of the steps
c) and d) may at first sight seem counterintuitive, since d) can be
clearly visualized as smoothing the vortex surface, whereas c) seems
more severe and does not constitute a smoothing step in the strict
sense; it can change the connectivity of the vortex world-sheets.
The reason c) is nevertheless carried out before d) lies in the fact
that including c) in the smoothing procedure in practice has little effect
on the observables measured here, while step d) is the chief source of
changes in the infrared properties of the surface ensemble, cf.
Fig.~\ref{fneu1}. The weak effect of step c) is due to the fact that
instances where c) is applied are rather rare.}
\label{smodef}
\end{figure}

Another way to remove the artificial ultraviolet fluctuations of the
center projection vortices is the smoothing procedure first discussed in
\cite{bertle}. It operates using elementary cube transformations of 
the type already introduced in section \ref{chilsec} in connection with 
the removal of topological ambiguities in the lattice vortex surface
configurations. The difference lies in the condition for accepting
an elementary cube transformation. The smoothing procedure is defined
by accepting such an elementary update whenever it implies a net
decrease in the number of vortex plaquettes, and it can be further
split up into a progression of steps characterized by the precise way
in which the update affects the vortex surface, as displayed in 
Fig.~\ref{smodef}. Note also that repeated smoothing sweeps through the
lattice are performed, until no further elementary cube transformations
of the type under consideration are possible; thus, smoothing can
propagate information over distances of more than one lattice spacing.

This smoothing procedure depletes the center projection vortex density while
keeping the long-range static quark potential largely intact, in accordance 
with the interpretation of the center projection vortices discussed above in 
connection with Fig.~\ref{flucpv}. However, in contrast to the blocking
procedure presented in the previous section, preservation of the
string tension is not an exact property of smoothing. Thus, invariance
of the static potential can be used as a criterion to determine how far 
the smoothing procedure can be applied before it begins to truncate 
relevant physical information about the confining thick vortex structures,
and should therefore be stopped.

\begin{figure}
\centerline{
\epsfysize=6.5cm
\epsffile{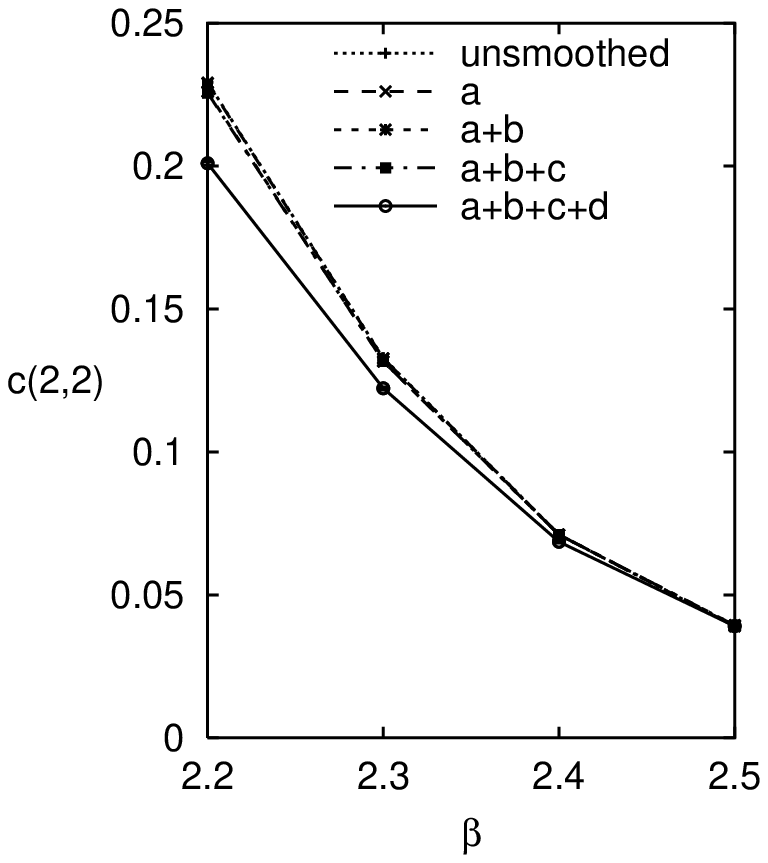}
\hspace{-0.3cm}
\epsfysize=6.5cm
\epsffile{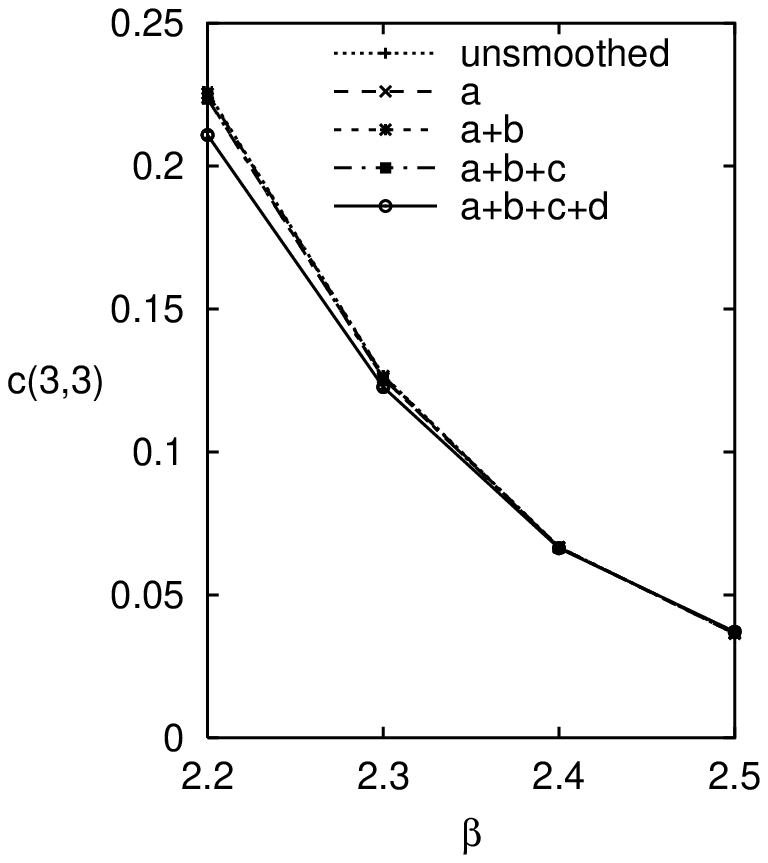}
\hspace{-0.3cm}
\epsfysize=6.5cm
\epsffile{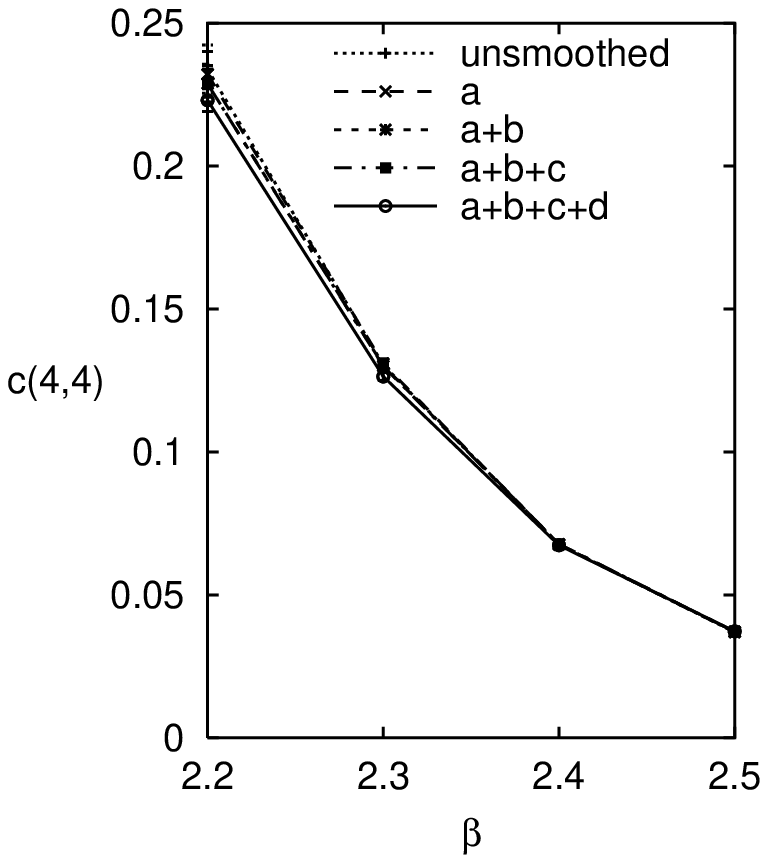}
\vspace{0.5cm}
}
\caption{\em Creutz ratios $c(2,2)$, $c(3,3)$ and $c(4,4)$ as a function of 
the inverse coupling $\beta $, for different versions of smoothing.
Incorporating smoothing steps a) through c), cf. Fig.~\ref{smodef}, leads
to no appreciable change in the Creutz ratios compared with the unsmoothed
ensemble. Only smoothing step d) has a non-negligible effect on the
static quark potential.}
\label{fneu1}
\end{figure}

The effect of the different smoothing steps on Creutz ratios is
displayed in Fig.~\ref{fneu1}. The Creutz ratios are clearly
unaffected by the smoothing steps a) through c), whereas a suppression
is seen after step d). At first sight, therefore, steps a) through c) only
remove spurious ultraviolet fluctuations from the vortex configurations,
whereas step d) begins to truncate relevant information. However, two
trends are visible in Fig.~\ref{fneu1} which deserve further
comment. For one, the suppression effect becomes weaker as one progresses
to larger Wilson loops. This is natural; a finite number of local smoothing
steps, which propagates information over a finite distance, cannot influence
correlations on scales larger than that distance\footnote{Note that an
analogous argument applies with regard to cooling steps applied to 
lattice Yang-Mills configurations.}. Thus, strictly speaking, the asymptotic
string tension is not modified by smoothing. However, already the medium
range behavior of the confining static quark potential constitutes
relevant nonperturbative information, e.g. through its influence on
hadronic properties. In this sense, smoothing step d) does truncate
important nonperturbative effects carried by the vortices.

On the other hand, according to Fig.~\ref{fneu1}, the
suppression of the Creutz ratios by smoothing step d) also weakens
as the inverse coupling $\beta $ is increased. This can be understood
from the fact that smoothing is not defined in a renormalization group
invariant manner. Elementary smoothing operations are defined locally on the
scale of one lattice spacing; as one increases $\beta $, this spacing
decreases. Therefore, some fluctuations of the vortex surfaces which occur
on the scale of one lattice spacing at low $\beta $ (and are therefore
removed by smoothing) remain unaffected by smoothing at higher $\beta $
because they then extend over more than one lattice 
spacing\footnote{Renormalization group invariance would presumably be
restored by considering smoothing steps of increasingly nonlocal nature
as $\beta $ is augmented.}. The trend visible in 
Fig.~\ref{fneu1} suggests that, at $\beta =2.5 $, one roughly
reaches the point where smoothing step d) just stops truncating relevant
physical information (and, presumably, still removes spurious ultraviolet
fluctuations). At higher $\beta $, smoothing step d) can be expected to
even leave some of these ultraviolet fluctuations intact.

As a consequence of the above discussion, the authors conclude that, at
$\beta =2.3 $, the topological susceptibility measured from center
projection vortices before applying smoothing step d) constitutes an
upper limit on the physical susceptibility carried by the confining thick 
physical vortices characterizing infrared Yang-Mills theory; the
measurement after smoothing step d) constitutes a lower limit. At
$\beta =2.5 $, on the other hand, the value measured after
smoothing step d) should give a rough indication of the 
aforementioned relevant physical thick vortex susceptibility. 
It will be seen that these characterizations at different $\beta $ 
are consistent with each other.

\section{Numerical Measurements and Discussion}
\label{ress}
Having presented all the elements employed in the analysis, the complete
procedure used to extract the topological susceptibility can be summarized 
as follows:

\begin{itemize}
\item
Generate an ensemble of $SU(2)$ lattice Yang-Mills configurations.
\item
Transform the configurations to the maximal center gauge and perform
center projection.
\item
Remove ultraviolet fluctuations of the center projection vortex surfaces
by either blocking or smoothing, cf. section \ref{eliart}.
\item
Randomly assign orientations to the dual lattice plaquettes making up the
surfaces, with a choice of bias which either maximizes or minimizes the
monopole line density, cf. the beginning of section \ref{chilsec}.
\item
Remove ambiguities in the vortex surfaces, i.e. lines along which
vortices intersect and monopole lines coinciding with singular surface
points, cf. section \ref{chilsec}.
\item
Evaluate the topological charge carried by the singular points,
cf. section \ref{chilsec}.
\end{itemize}

\begin{figure}
\centerline{
\hspace{0.4cm}
\epsfysize=5.5cm
\epsffile{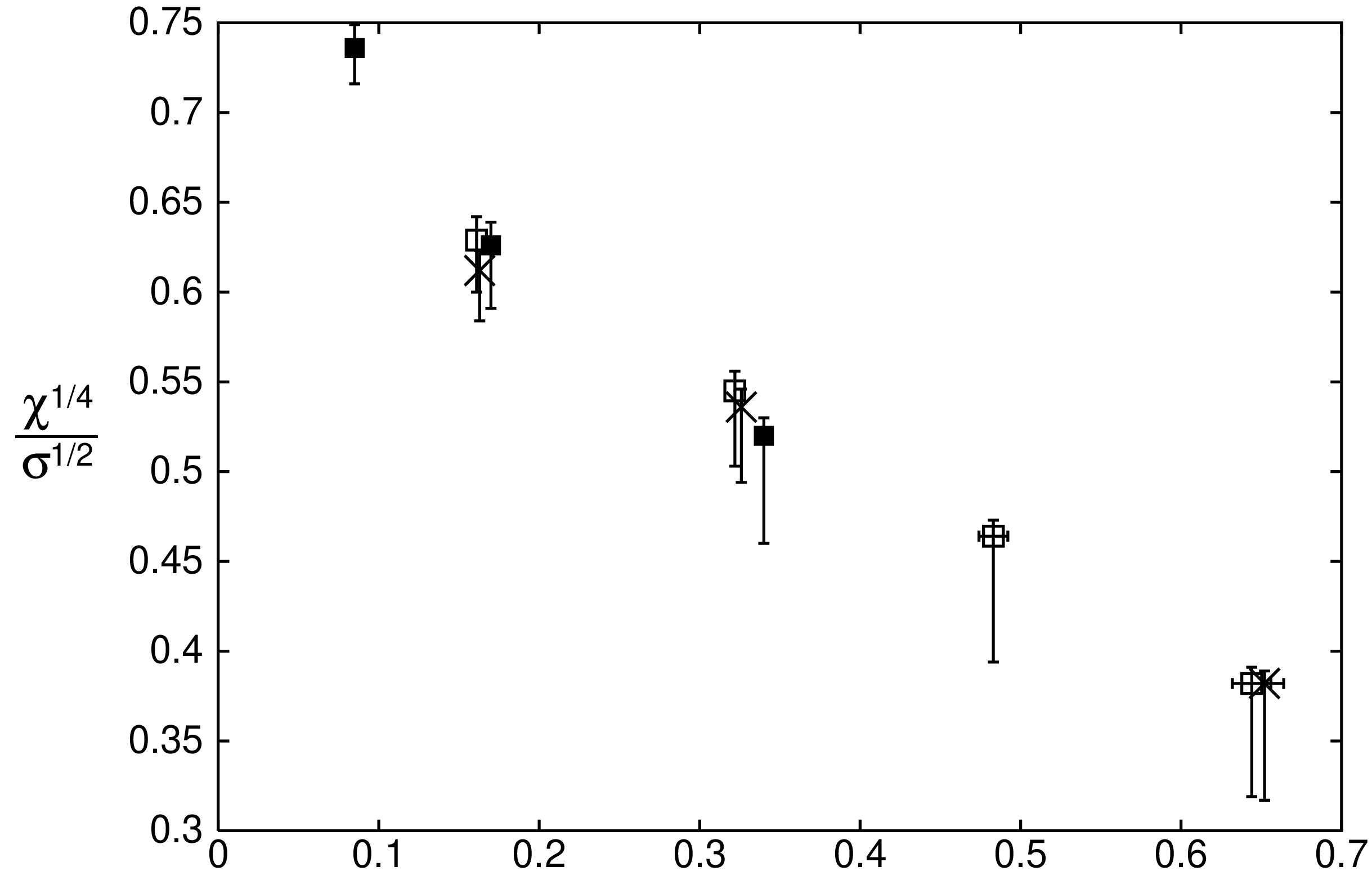}
\hspace{0.1cm}
\epsfysize=5.5cm
\epsffile{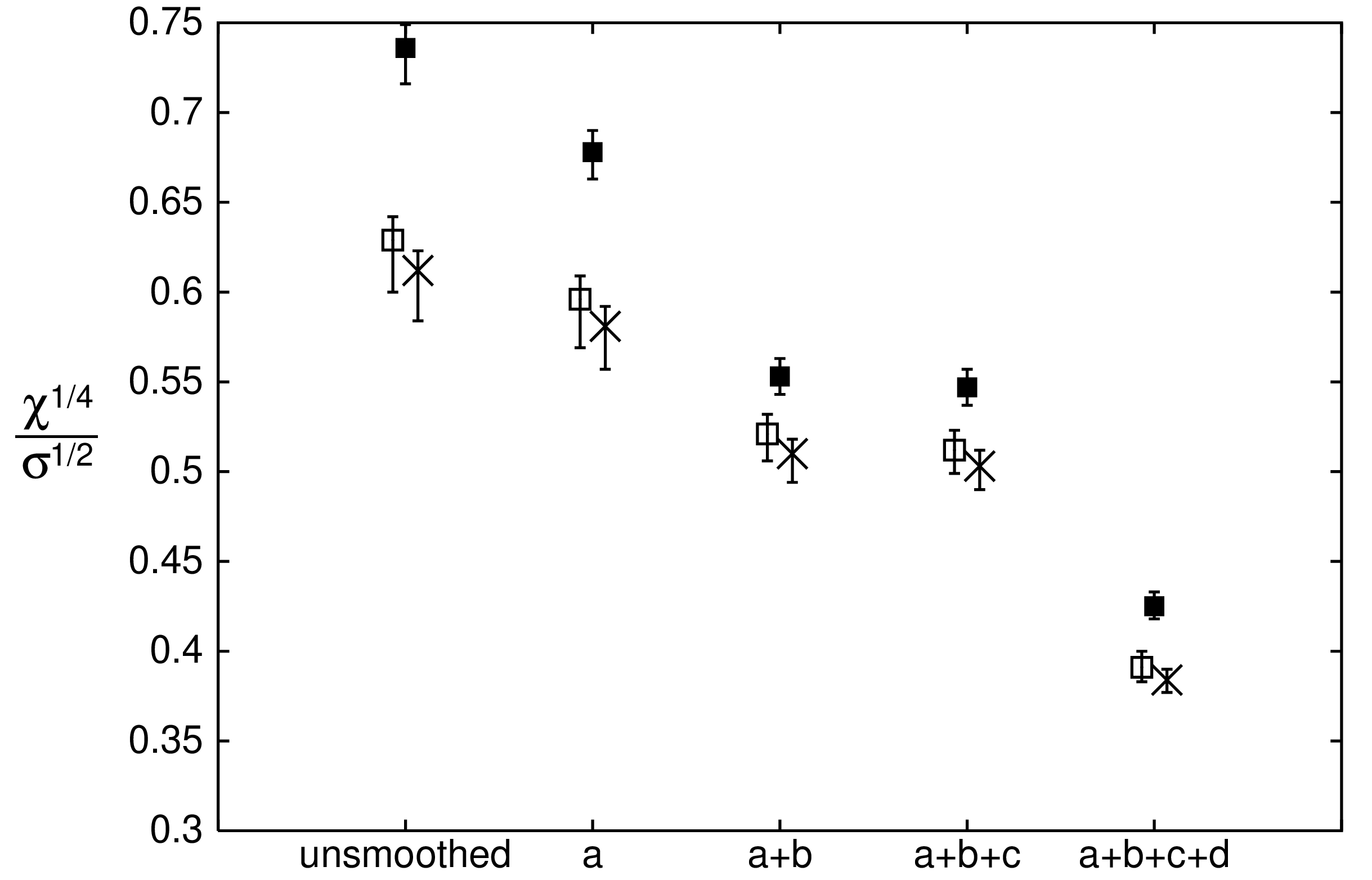}
\vspace{0.5cm}
}
\caption{\em Fourth root of the topological susceptibility $\chi $ carried
by center projection vortices, in units of the square root of the string
tension $\sigma $. Measurements are shown as a function of the blocking
scale $a$ (left) and of the smoothing steps (right). Open squares correspond
to $\beta =2.3$ on a $12^4 $ lattice, crosses to $\beta =2.3$ on a $16^4 $
lattice, and filled squares to $\beta =2.5$ on a $16^4 $ lattice. Error
bars are discussed in the main text.}
\label{resub}
\end{figure}

Fig.~\ref{resub} depicts the results for the topological susceptibility 
$\chi $ of the different center projection vortex ensembles considered. 
Measurements are displayed as a function of the blocking scale (i.e. the 
spacing of the blocked lattice), and as a function of the smoothing steps. 
The fact that the $\beta=2.3 $ and the $\beta=2.5 $ values in the right-hand 
panel in Fig.~\ref{resub} do not lie on a universal curve (as opposed to 
the left-hand panel) is natural, since the smoothing steps defining the 
horizontal axis are not constructed in a renormalization group invariant 
manner, cf. the discussion in the previous section.

The vertical error bars in Fig.~\ref{resub} are compounded from three 
sources: The statistical uncertainty of the susceptibility measurement, the
statistical uncertainty of the string tension measurement, and a systematical
uncertainty stemming from the vortex surface ambiguity removal procedure
discussed in section \ref{chilsec}. The latter was estimated as follows:
The ambiguity removal procedure systematically increases the vortex surface
density; the densities before and after this procedure were recorded.
To obtain a measure for the uncertainty introduced by this alteration
of the surface configurations, the variation of the topological 
susceptibility was determined which would result from readjusting the 
physical scale such as to equate the aforementioned densities. Under such a 
rescaling, the topological susceptibility varies as the square of the vortex 
density, in view of the dimensions of the two quantities. The variation of 
the susceptibility extracted in this way was compounded only into the
downward uncertainty of the measurements displayed in Fig.~\ref{resub},
since surface ambiguity removal systematically increases the vortex density. 
Consequently, as is markedly visible in the left-hand panel in 
Fig.~\ref{resub}, the downward error bar is always larger than the upward 
error bar. Evidently, for blocked configurations, the systematical 
uncertainty dominates over the statistical one. Presumably, the coarse 
lattice, onto which the surface configurations are forced by blocking, does 
not allow the vortices to avoid one another, thereby inducing a large 
density of lines on which vortex surfaces intersect. These ambiguities 
subsequently have to be removed again by correspondingly abundant 
applications of elementary cube transformations, as described in section 
\ref{chilsec}. The strong alteration of the surfaces implied by this leads 
to a large systematical uncertainty of the type discussed above. In marked 
contrast to this, in the case of smoothing, cf. Fig.~\ref{resub} (right), 
the difference between the upward and the downward uncertainties is not 
appreciable. Smoothing thus seems to constitute a good preconditioner for 
the vortex surface ambiguity removal procedure.

In addition to these uncertainties in $\chi /\sigma^{2} $, the left-hand
panel in Fig.~\ref{resub} also displays horizontal error bars stemming from 
the statistical error in the lattice measurement of $\sigma a^2 $, which 
was used to determine the lattice spacing $a$ by equating 
$\sqrt{\sigma } =440$ MeV.

\begin{figure}
\centerline{
\hspace{0.8cm}
\epsfysize=5.5cm
\epsffile{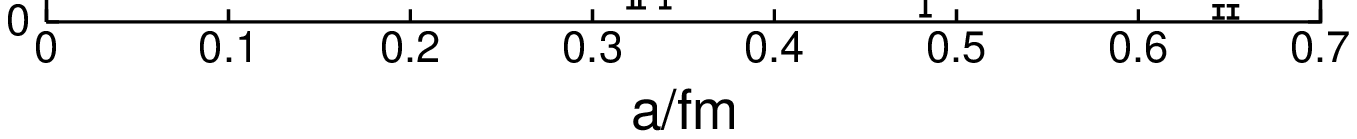}
\hspace{0.9cm}
\epsfysize=5.5cm
\epsffile{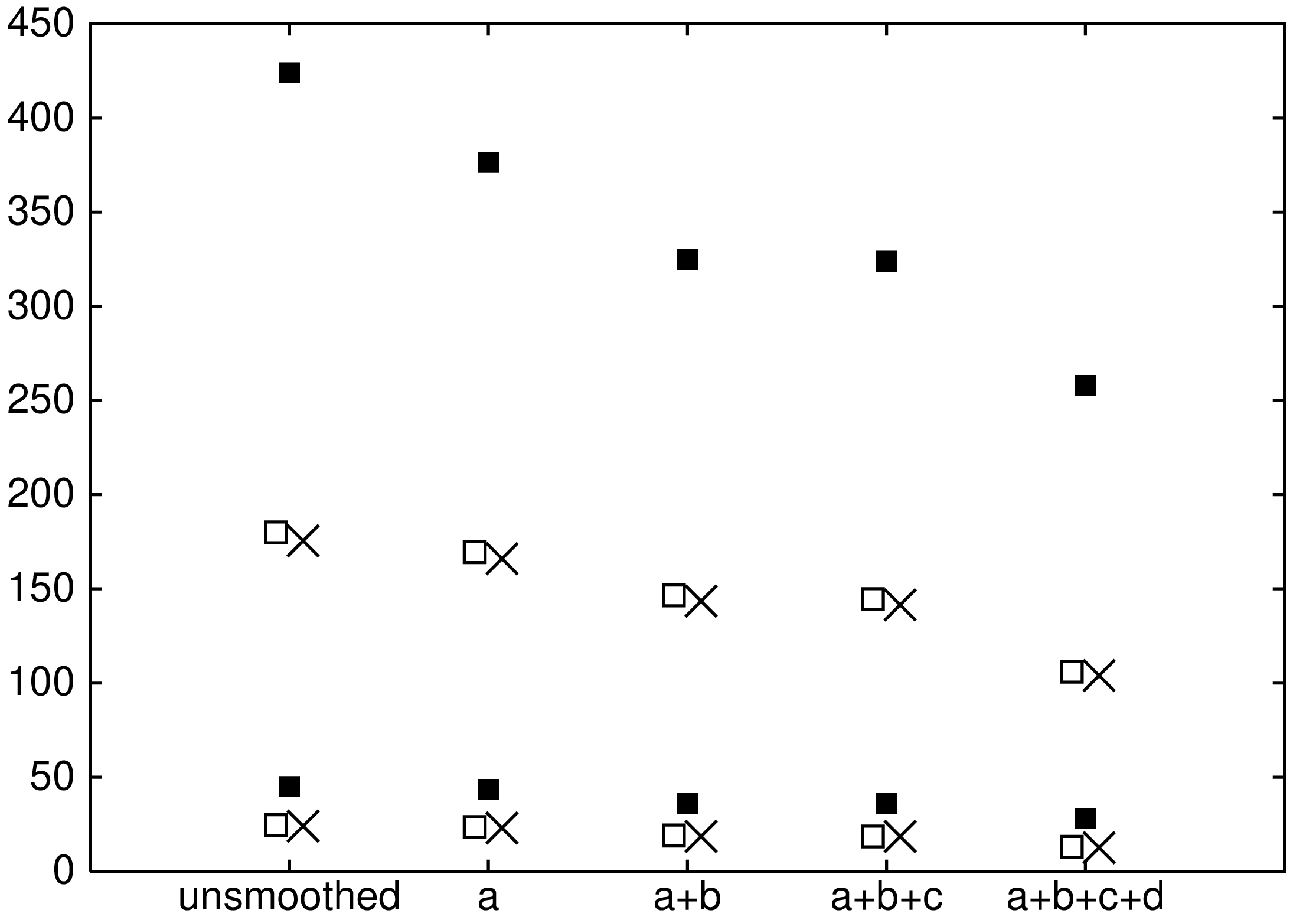}
\vspace{0.5cm}
}
\caption{\em Monopole line density $\rho $ as a function of the blocking 
scale (left) and of the smoothing step (right). For each point on the 
horizontal axes, both the maximal and the minimal
densities reached by the vortex plaquette reorientation procedure of
section \ref{chilsec} are shown, where in the right-hand panel, open
squares correspond to $\beta =2.3$ on a $12^4 $ lattice, crosses to
$\beta =2.3$ on a $16^4 $ lattice, and filled squares to $\beta =2.5$ on 
a $16^4 $ lattice. In the left-hand panel, identification of the different
$\beta $ values and lattices is foregone for the sake of legibility; they
can however be inferred by comparing with Fig.~\ref{resub} (left), since
the ordering of the data in the blocking scale is identical. Instead, the
data are represented by vertical bars to indicate the rise in the densities
induced by the vortex surface ambiguity removal procedure of section
\ref{chilsec}. Thus, the lower end of each vertical bar gives the monopole
density originally defined by the plaquette reorientation procedure, 
whereas the upper end of each bar represents the density after the 
subsequent ambiguity removal; this is therefore the density at which the
Pontryagin index was ultimately measured. The inset is simply an 
enlargement of the range $[0.3\, \mbox{\small fm},0.7\, \mbox{\small fm}]$ 
in the blocking scale. In the case of smoothing (right), the variation
of the monopole density through the vortex surface ambiguity removal
is always smaller than the symbols displayed; in fact, in marked contrast
to the case of blocking, this variation becomes negligible with progressive 
smoothing steps, cf. also the discussion of the systematic uncertainty in 
Fig.~\ref{resub}.}
\label{monub}
\end{figure}

Before proceeding to interpret Fig.~\ref{resub}, a
discussion of the monopole density dependence of the results is in order.
The measurements displayed in Fig.~\ref{resub} were obtained using
the maximal monopole density reached via the biased vortex plaquette
reorientation procedure described at the beginning of section \ref{chilsec}.
If one conversely minimizes the monopole density, the values shown in
Fig.~\ref{resub} only vary by at most 1\%, i.e. by
considerably less than the uncertainty of the measurement. Thus, for
practical purposes, the topological susceptibility is independent of
the monopole density. To illustrate the significance of this result,
Fig.~\ref{monub} displays the aforementioned maximal
and minimal monopole line densities considered for each measured data
point in Fig.~\ref{resub}. For comparison, the zero-temperature
monopole line density measured in full $SU(2)$ Yang-Mills theory in the
maximal Abelian gauge \cite{borny} amounts to
$\rho_{mag} = 64/\mbox{fm}^{3} $.

The phenomenon that the topological susceptibility is independent of the
monopole density has been observed before in the random vortex surface
model \cite{preptop}. Also the reasons for this independence there and
here are similar. Most importantly, the dominant proportion of the 
topological charge is carried by so-called writhing points of the
vortex surfaces as opposed to intersection points in the usual sense.
The former class of singular points is distinguished from the latter
as follows: At intersection points, two distinct surface segments share 
one point, but one cannot reach one surface segment from the other by 
proceeding along plaquettes which share a link. Writhing points on the 
other hand are characterized precisely by the opposite; all plaquettes 
attached to such a point can be connected by proceeding along plaquettes 
which share a link. In this sense, there is only one surface segment at 
a writhing point. As a consequence of this structure, the associated 
contribution to the Pontryagin index is manifestly invariant under changes
of the monopole configuration: Encircling a writhing point by a monopole 
loop implies inverting the orientations of all plaquettes attached to the 
point, since they are all connected via links. As a result, all pairs of 
orthogonal plaquettes retain their relative orientation, and the Pontryagin 
index is unchanged \cite{preptop}.

\begin{figure}
\centerline{
\hspace{0.4cm}
\epsfysize=5.5cm
\epsffile{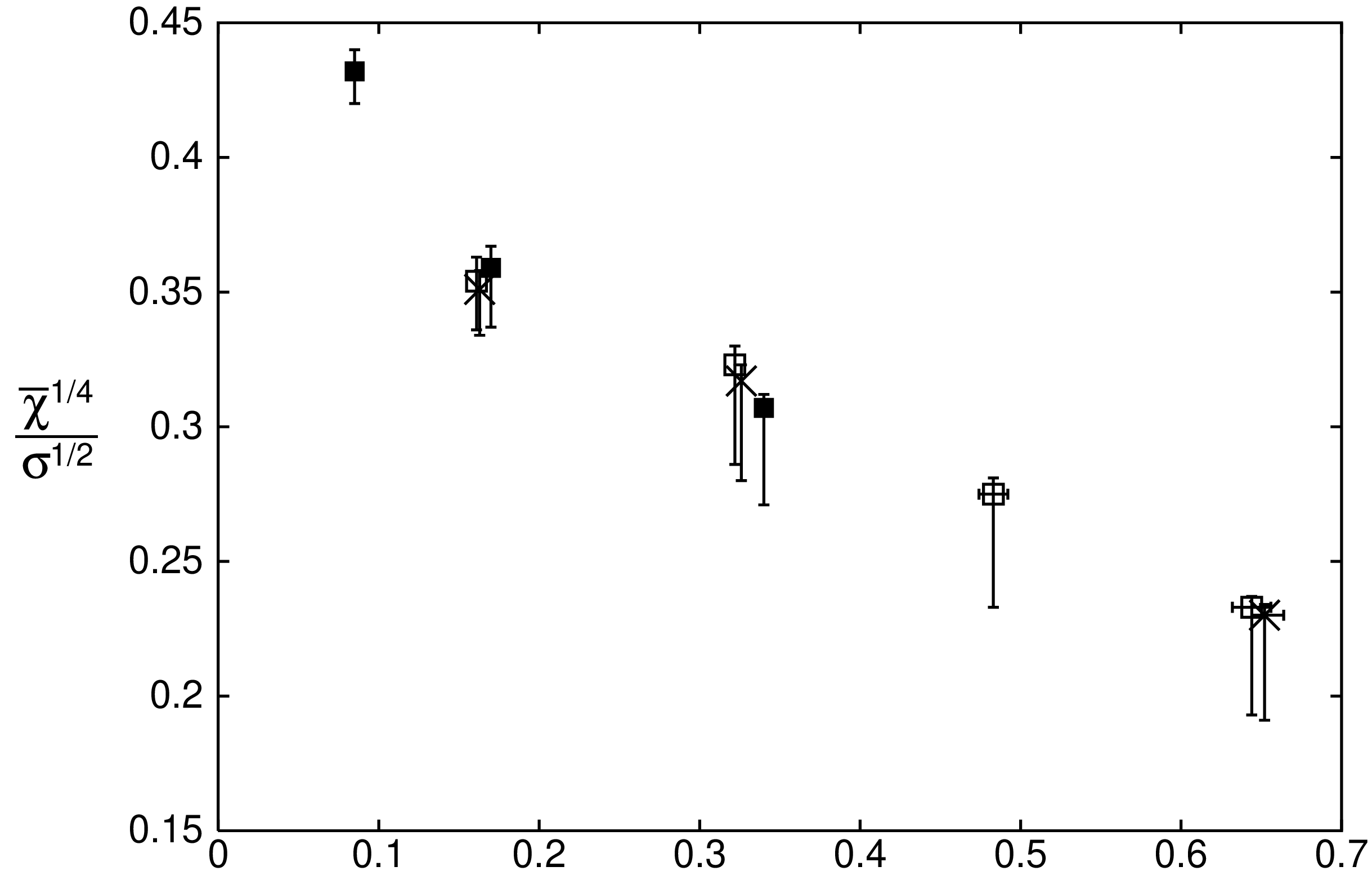}
\hspace{0.1cm}
\epsfysize=5.5cm
\epsffile{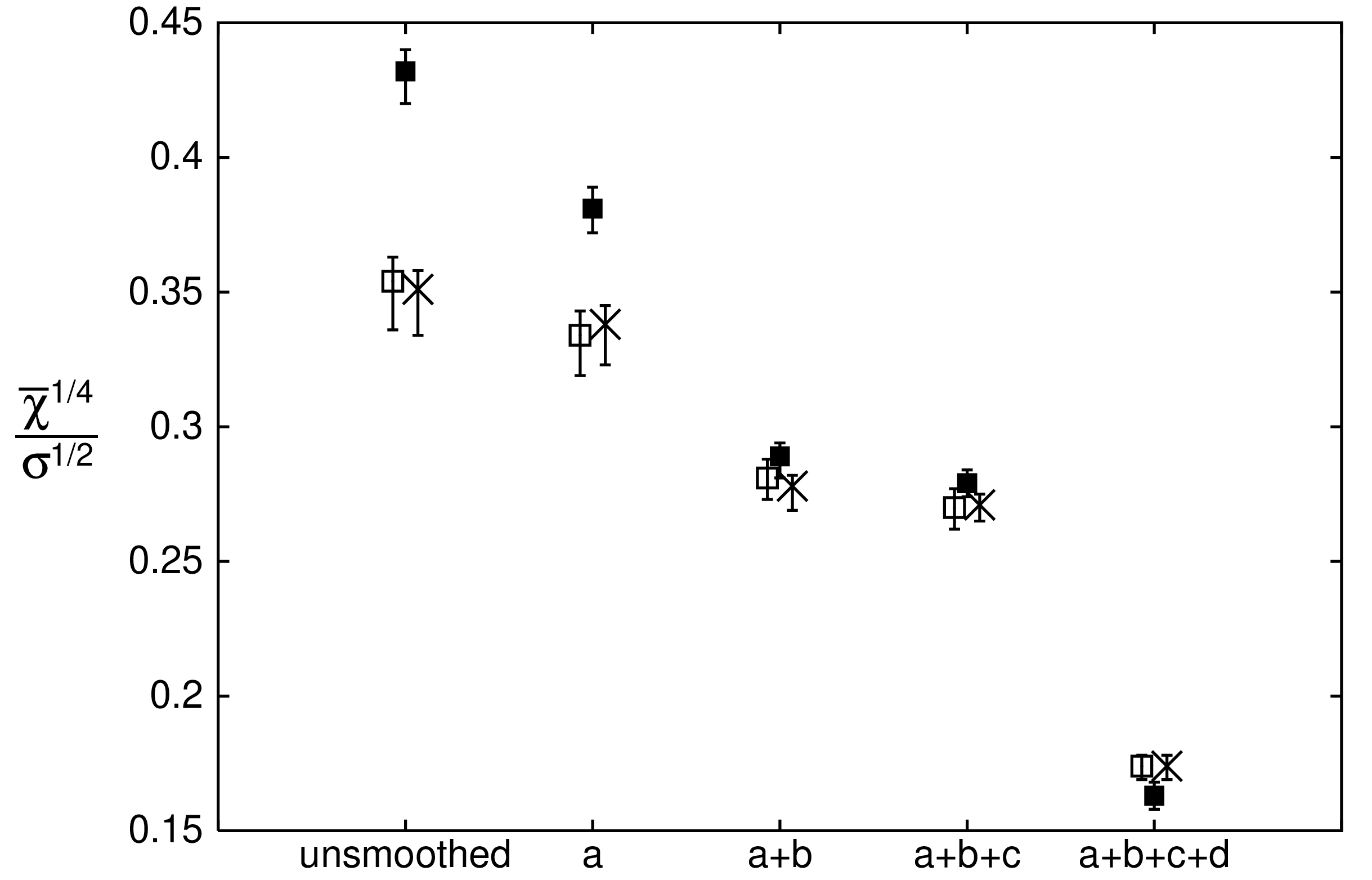}
\vspace{0.5cm}
}
\caption{\em Truncated topological susceptibility $\bar{\chi } $ of center
projection vortices obtained by disregarding writhing points and taking only
contributions from intersection points into account. For legibility, the
fourth root of the susceptibility is displayed, in units of the square
root of the string tension $\sigma $. Measurements are shown as a function 
of the blocking scale $a$ (left) and of the smoothing steps (right). Open 
squares correspond to $\beta =2.3$ on a $12^4 $ lattice, crosses to 
$\beta =2.3$ on a $16^4 $ lattice, and filled squares to $\beta =2.5$ on 
a $16^4 $ lattice.}
\label{risub}
\end{figure}

The topological susceptibility therefore must be independent of the
monopole density to the extent that it is dominated by the contributions
from writhing points. To corroborate this dominance, Fig.~\ref{risub} 
displays the topological susceptibility $\bar{\chi } $ obtained by 
discarding writhing points (in all other respects, Fig.~\ref{risub}
is completely analogous to Fig.~\ref{resub}).
Evidently, the fourth root of this truncated susceptibility
is only roughly half as large as the fourth root of the full one; i.e.,
the contribution of intersection points to the full topological
susceptibility is suppressed compared with the contribution from
writhing points by roughly a factor $2^4 $.

Note furthermore that Fig.~\ref{risub} was again obtained
using the maximal monopole density reached via the biased vortex plaquette
reorientation procedure of section \ref{chilsec}. If one instead uses
the minimal monopole density, the variation of the results in
Fig.~\ref{risub} still is rather weak; it amounts to no more than 5\%, 
which is comparable to the statistical uncertainty of the
measurement. This at first sight surprisingly weak dependence is
presumably due to the high degree of non-orientability of the vortex
surfaces. This non-orientability enforces a certain minimal monopole
density which cannot be removed by the aforementioned vortex reorientation
procedure. Evidently already this minimal density suffices to randomize
the signs of the intersection point contributions to the Pontryagin index
to such an extent that additional random changes of the signs, induced
by adding monopole loops on the vortex surfaces, do not strongly influence
the associated topological susceptibility.

The fact that the topological susceptibility is virtually independent of
the monopole density allows to predict the former without going to the
trouble of explicitly determining the monopole content of each lattice
Yang-Mills configuration considered. The remaining task lies in extracting
from Fig.~\ref{resub} the physical value of the topological 
susceptibility obtained after eliminating spurious ultraviolet fluctuations 
of the center projection vortex surfaces. Starting with the left-hand panel
in Fig.~\ref{resub}, the discussion in section \ref{blosec} led to the 
conclusion that the residual susceptibility $\chi $ at a blocking scale of 
0.4 fm represents an upper limit for the physical susceptibility 
$\chi_{phys} $, whereas the value at 0.6 fm constitutes a lower limit. 
In the extreme cases admitted by the error bars, this implies
\begin{equation}
(150\, \mbox{MeV} )^4 \leq \chi_{phys} \leq (224\, \mbox{MeV} )^4 \ ,
\label{estib}
\end{equation}
where $\sqrt{\sigma } =440$ MeV was used. Likewise, in the right-hand panel
in Fig.~\ref{resub}, the $\beta=2.3 $ data extracted using smoothing steps 
a) through c) limit the physical susceptibility from above, whereas the 
$\beta=2.3 $ data obtained using smoothing steps a) through d) limit it 
from below, cf. the discussion in section \ref{smosec}. Therefore, one has
in the extreme cases admitted by the error bars
\begin{equation}
(166\, \mbox{MeV} )^4 \leq \chi_{phys} \leq (230\, \mbox{MeV} )^4 \ .
\label{estis}
\end{equation}
This is furthermore consistent with the value obtained at $\beta=2.5 $
using smoothing steps a) through d), namely $\chi^{1/4} = (187\pm 3)$ MeV
(only statistical error quoted).

In summary, the results for the topological susceptibility carried by the 
physical thick vortex content of lattice Yang-Mills configurations, as 
estimated within the different schemes of eliminating spurious ultraviolet 
fluctuations of the associated thin center projection vortices, are consistent
with one another. While considerable systematic uncertainties are inherent 
in all these determinations, they correspond well with values extracted 
from the full $SU(2)$ lattice Yang-Mills configurations \cite{stama}. 
The latter values are located roughly at the center of the range admitted 
by (\ref{estib}) and (\ref{estis}). This suggests that the topological 
properties of the Yang-Mills ensemble can be accounted for in terms of 
the vortex content of the gauge field configurations, just as is the case
for the confining properties. The vortex picture appears suited to provide 
a unified description of these two different nonperturbative aspects of 
Yang-Mills theory.

\section*{Acknowledgments}
M.E. acknowledges informative discussions with K.~Langfeld and, especially,
H.~Reinhardt, in particular concerning their results in \cite{kurt} together
with A.~Sch\"afke, prior to publication of that work. M.E. furthermore 
acknowledges DFG financial support under grant DFG En 415/1-1, including 
the funding of a collaborative stay at TU Wien; also the hospitality of the 
Institut f\"ur Kernphysik there is acknowledged. R.B. reciprocally is grateful 
for the hospitality during a stay at the Institut f\"ur Theoretische 
Physik at T\"ubingen, funded by DFG under grant DFG Al 279/3-3, in the
course of which this work was finalized. M.F. and R.B. are supported by 
Fonds zur F\"orderung der Wissenschaftlichen Forschung under P13997-TPH.

\end{document}